\documentclass[sn-mathphys-num]{sn-jnl}

\usepackage{graphicx}%
\usepackage{multirow}%
\usepackage{amsmath,amssymb,amsfonts}%
\usepackage{amsthm}%
\usepackage{mathrsfs}%
\usepackage[title]{appendix}%
\usepackage{xcolor}%
\usepackage{textcomp}%
\usepackage{manyfoot}%
\usepackage{booktabs}%
\usepackage{algorithm}%
\usepackage{algorithmicx}%
\usepackage{algpseudocode}%
\usepackage{listings}%

\theoremstyle{thmstyleone}%
\theoremstyle{thmstyletwo}%

\theoremstyle{thmstylethree}%

\begin{document}

\title[Article Title]{CADMR: Cross-Attention and Disentangled Learning for Multimodal Recommender Systems}


\author*[1,2]{\fnm{Yasser} \sur{KHALAFAOUI}}\email{mykhalafaoui@alteca.fr}

\author[1]{\fnm{Martino} \sur{LOVISETTO}}\email{mlovisetto@alteca.fr}

\author[3]{\fnm{Basarab} \sur{MATEI}}\email{basarab.matei@lipn.univ-paris13.fr}

\author[2]{\fnm{Nistor} \sur{GROZAVU}}\email{nistor.grozavu@cyu.fr}

\affil*[1]{\orgdiv{R\&D Dept.}, \orgname{ALTECA}, \orgaddress{ \city{Lyon}, \country{France}}}

\affil[2]{\orgdiv{CS Dept.}, \orgname{CY Cergy Paris University}, \orgaddress{\city{Pontoise}, \country{France}}}

\affil[3]{\orgdiv{CS Dept.}, \orgname{Sorbonne Paris Nord University}, \orgaddress{\city{Villetaneuse}, \country{France}}}


\abstract{The increasing availability and diversity of multimodal data in recommender systems offer new avenues for enhancing recommendation accuracy and user satisfaction. However, these systems must contend with high-dimensional, sparse user-item rating matrices, where reconstructing the matrix with only small subsets of preferred items for each user poses a significant challenge. To address this, we propose CADMR, a novel autoencoder-based multimodal recommender system framework. CADMR leverages multi-head cross-attention mechanisms and Disentangled Learning to effectively integrate and utilize heterogeneous multimodal data in reconstructing the rating matrix. Our approach first disentangles modality-specific features while preserving their interdependence, thereby learning a joint latent representation. The multi-head cross-attention mechanism is then applied to enhance user-item interaction representations with respect to the learned multimodal item latent representations. We evaluate CADMR on three benchmark datasets, demonstrating significant performance improvements over state-of-the-art methods.}

\keywords{Recommender systems, Multimodal recommendation, Multimedia and multimodal retrieval, Attention, Disentangled learning}



\maketitle

\section{Introduction}\label{sec1}

Recommender systems have become a cornerstone of modern digital experiences, playing a critical role in various domains such as streaming services and social media. By filtering and presenting relevant content from an overwhelming array of options, these systems enhance user satisfaction and engagement. Recommender systems can be classified into three main groups: content-based, collaborative filtering (CF), and hybrid algorithms \cite{roy2022systematic}. Content-based filtering uses item features to recommend items similar to those a user has liked based on previous actions or explicit feedback. This approach assumes that users are likely to choose items with features similar to those they have previously liked. However, this method struggles when recommending new items with novel features. On the other hand, collaborative filtering predicts a user’s interests based on their historical behavior and the preferences of a large number of other users \cite{han2021glocal}. Hybrid approaches combine techniques from both collaborative filtering and content-based filtering to achieve more accurate results \cite{zhou2023comprehensive}.\par

CF remains one of the most widely utilized approaches for recommendation tasks. However, it is limited by the sparsity of the user-item rating matrix and the cold start problem, where new users lack historical behavior data. Various approaches have been proposed to overcome these limitations. Some methods \cite{hernando2016non,zhang2005using} leverage dimensionality reduction and clustering to mitigate the effects of sparsity. These approaches aim to use a low-dimensional latent representation of the rating matrix instead of the original sparse matrix to estimate user-item interactions. Common approaches include Matrix Factorization and Singular Value Decomposition \cite{roy2022systematic}. Other methods employ autoencoders (AE) to learn the low-dimensional feature space of a given sparse rating matrix. AEs are particularly useful for capturing non-linear relationships and encoding complex features into a compact latent representation. For instance, I-AutoRec \cite{sedhain2015autorec} utilizes an item-based AE to project high-dimensional matrix entries into a low-dimensional latent hidden space, reconstructing the entries in the output space to predict missing ratings. GLocal-K \cite{han2021glocal} presents a two-stage framework that pre-trains the autoencoder using a local kernel matrix and then fine-tunes the model using a global convolution kernel. However, these approaches primarily focus on the rating matrix and often overlook the potential of incorporating side information.\par

Recently, multimodal-based recommender systems have gained attention, leveraging diverse data sources to improve recommendations. Zhang et al. \cite{zhang2017joint} propose a method that uses a Multi-Layer Perceptron (MLP) to project multimodal data into a unified latent space, with the learned features transferred to user and item embeddings, which are optimized using a reconstruction loss. Similarly, Zheng et al. \cite{zheng2017joint} introduce DeepCoNN, where text review embeddings are fed into two parallel CNN architectures to learn both user and item representations. Despite these advances, existing multimodal approaches typically learn user and item embeddings separately, neglecting the crucial relationship between the user-item rating matrix and the various data modalities. This oversight is significant because user-item interactions are inherently influenced by multiple factors present in multimodal data, which should be integrated into the recommendation process. However, current methods fail to leverage these rich interactions, potentially limiting their effectiveness in capturing the full complexity of user preferences.\par

To address the limitations of existing approaches, this paper introduces the framework CADMR (Cross-Attention and Disentangled Learning for Multimodal Recommender Systems). It begins with a pretraining phase where modality-specific features are disentangled and projected into a shared latent space. Simultaneously, the user-item rating matrix is processed through an autoencoder to capture a compact representation of user-item interactions. During the fine-tuning phase, a multi-head cross-attention mechanism is employed to integrate the rating matrix with the multimodal latent representations. This mechanism allows each user-item interaction to attend to the most relevant complementary information from the multimodal data. Finally, the enhanced rating matrix is reintroduced into the autoencoder for the reconstruction of the final, refined rating matrix. Our contributions can be summarized as follows:

\begin{itemize} 
    \item We propose CADMR, a multimodal recommender system framework incorporating disentangled learning and cross-attention. 
    \item We introduce a novel approach for integrating multimodal information into the reconstruction of the rating matrix using a cross-attention mechanism, ensuring that user-item interactions are informed by the most relevant multimodal features. 
    \item We conduct extensive experiments on benchmark datasets, validating the effectiveness and efficiency of CADMR. The results demonstrate superior performance compared to other state-of-the-art methods. 
\end{itemize}

\section{Related Work}
\subsection{Multimodal Collaborative Filtering}
The multimodal collaborative filtering-based recommender system aims to learn informative representations of users and items by leveraging multimodal features. He et al. \cite{he2016vbpr} extracted visual features using a CNN, which they concatenated with ID embeddings to produce a refined item representation, which is then fed to a scalable factorization model to learn users' opinions. Joint Representation Learning for top-n recommendations (JRL) \cite{zhang2017joint} leverages the MLP architecture to learn a unified latent representation for the multimodal data. A mapping matrix is then learned to transfer the unified multimodal representation into user and item embeddings, while the latter are learned using a reconstruction loss. Kang et al. \cite{kang2017visually} build on existing BPR recommender systems by introducing a Siamese CNN where the last layer of the CNN produces the item representation. Moreover, the framework learns pixel-level representations by training the image representation and the recommender system jointly. While these methods successfully incorporate multimodal features into collaborative filtering frameworks, they often suffer from several limitations. One significant weakness is that these approaches typically treat multimodal data as supplementary to the core user-item interaction, rather than as integral components of the interaction itself. This can lead to suboptimal utilization of the rich information contained in the various modalities.

\subsection{Attention Networks}
Traditional multimodal recommender systems often struggle to effectively integrate and utilize multimodal information. The attention mechanism offers a solution by allowing the recommender system to selectively focus on different aspects of the multimodal data, thereby better capturing user preferences. For instance, ACF \cite{chen2017attentive} introduces a framework that incorporates both item-level and component-level attention, enabling the system to attend to high-level information (e.g., photos, videos) as well as finer details, such as specific regions of an image or frames within a video. Similarly, Disentangled Multimodal Representation Learning for Recommendation (DMRL) \cite{wang2022disentangled} takes a novel approach by dividing each modality into equal factors and learning a disentangled representation of these different modality feature factors. The attention mechanism is then applied to each factor to effectively model user preferences. UVCAN \cite{liu2019user} leverages the attention mechanism in micro-video recommendations by learning multimodal representations for both users and micro-videos through a co-attention mechanism between items and micro-videos. Additionally, it refines video representations by using them as input queries to capture user attention through multi-step reasoning.\par
Building upon existing approaches that utilize attention mechanisms, our method introduces a novel strategy starting with a pretraining phase where disentangled representations for the different modalities are learned, and the user-item rating matrix is reconstructed via an autoencoder. Following this, we apply a cross-attention mechanism between the user-item rating matrix and the learned multimodal representation. The multimodal-enhanced rating matrix is then projected back into the autoencoder for fine-tuning, ensuring a more integrated and robust modeling of user preferences.

\section{Methodology}
In this section, we present CADMR, a novel framework for multimodal recommender systems. Our approach advances existing methods by integrating cross-attention between the rating matrix and multimodal representations. We begin by detailing the learning process for disentangled multimodal representations. Next, we delve into the cross-attention mechanism. Finally, we describe the complete pipeline, from pre-training to the final reconstruction of the rating matrix.\par
\begin{figure}[t]
    \centering
    \includegraphics[width=\linewidth]{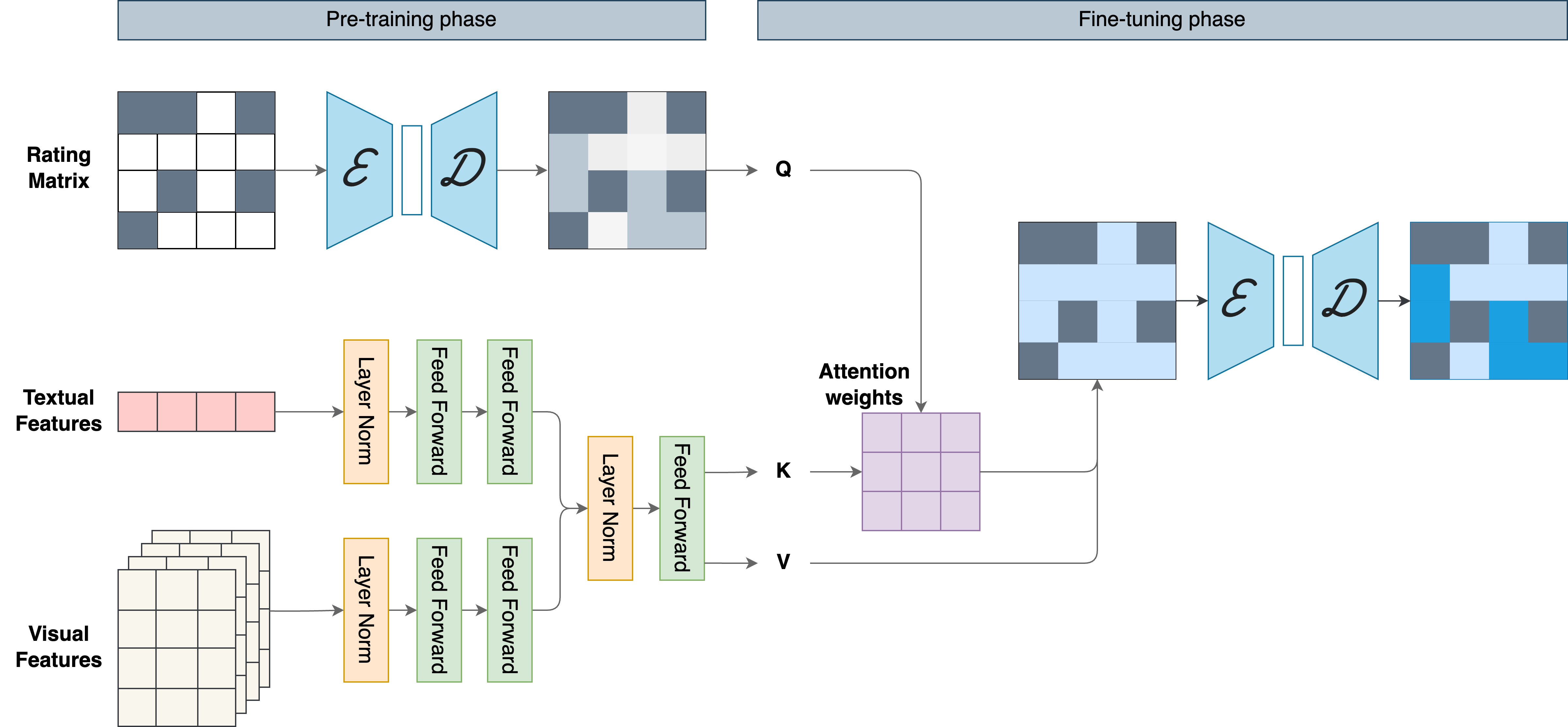}
    \caption{CADMR architecture for a multimodal recommender system. In the pretraining phase, the autoencoder and modality-specific feature extractors are trained separately. In the fine-tuning phase, a cross-attention mechanism integrates the user-item rating matrix with the unified multimodal representation, refining the matrix, which is then processed through the trained autoencoder to produce the final reconstructed rating matrix.}
    \label{fig:cadmr_architecture}
\end{figure}

Fig. \ref{fig:cadmr_architecture} illustrates the CADMR architecture for a multimodal recommender system, which operates in two distinct phases: pretraining and fine-tuning. During the pretraining phase, an autoencoder is trained on the user-item rating matrix while modality-specific feature extractors learn disentangled representations for textual and visual features. These extracted features undergo layer normalization and are processed through feedforward networks, in order to learn the unified multimodal representation. In the fine-tuning phase, a multi-head cross-attention mechanism is employed, allowing the user-item rating matrix, the query vector, to interact with the unified multimodal representation—treated as key, and value vectors—to generate attention weights. These weights are used to refine the rating matrix, which is then reintroduced into the autoencoder for final reconstruction, resulting in a more accurate and refined rating matrix.

\subsection{Preliminaries}
In what follows, we denote $\mathcal{R}\in \mathbb{R}^{I\times U}$ as the user-item rating or interaction matrix, with $I$ and $U$ representing the sets of items and users, respectively. We consider a user $u$ to have interacted with an item $i$,  if the corresponding entry $r_{i,u} \in \mathcal{R}$ is nonzero, otherwise, there was no interaction, and $r_{i,u} = 0$. In our setting, each user and item have a unique ID, and we assume that each item is associated with a set of multimodal information, such as item descriptions and images. Unlike existing approaches that treat the item ID as an independent modality, we chose to exclude it from our analysis, focusing instead on textual and visual data. These modalities provide richer, more descriptive features that are sufficient to capture the nuanced preferences of users and the intrinsic characteristics of items.

\subsection{Feature Extraction} 
The extracted multimodal features are fed separately into their respective two-layer neural networks to learn task-specific features relevant to the recommendation. We introduce a layer normalization before each of these neural networks to stabilize the learning process, improve the convergence speed, and ensure that the features fed into the networks are on a consistent scale. The projection neural network is defined as,
\begin{equation}\label{eq:projection_network}
    h  = f(W_1g(W_0N(x) + b_0) + b_1),
\end{equation}
where $x$ is the multimodal data features, $f$ and $g$ are the activation functions, $W_0, W_1$ and $b_0, b_1$ the corresponding layer weights and biases, and $N(.)$ is the layer normalization. It should be noted that the output $h = \{h_t, h_v\}$ can represent either the visual or textual feature representation.

\subsection{Disentangled Representation Learning}
Let $h_t\in \mathbb{R}^{D_t}$ and $h_v\in \mathbb{R}^{D_v}$ represent the textual and visual feature embeddings obtained using the feature extraction module, respectively, where $D_t$ and $D_v$ denote the dimensions of these feature spaces.\par
One of the main challenges in multimodal learning is that features extracted from different modalities often contain redundant or entangled information. This means that certain dimensions of the latent space might capture overlapping or correlated features between modalities, leading to inefficiencies and potential overfitting in downstream tasks. To address this issue, we employ disentangled learning, which aims to ensure that each dimension of the latent representations captures distinct and non-overlapping factors of variation. By disentangling the features, we improve the generalization capability of the model and reduce the risk of redundancy.\par
The disentangled loss is applied to both $h_t$ and $h_v$, encouraging each dimension of the latent representations to capture distinct, non-overlapping information. This is achieved through a Total Correlation (TC) loss, which measures the dependency along the dimensions of the latent representation. The TC loss can be expressed as,
\begin{equation}\label{eq:tc_loss}
    \mathcal{L}_{TC}(h) = \sum_{d=1}^{D}\mathbb{E}\left [log\frac{p(h^d)}{p(h^d|h^{d'})} \right ],
\end{equation}
where $p(h^d)$ is the marginal probability of the $d$-th dimension of the latent variable $h$, and $p(h^d|h^{d'})$ is the conditional probability of $h^d$ given the other dimensions $h^{d'}$.\par
By minimizing this loss, the learned representations $h_t$ and $h_v$ are encouraged to have minimal redundancy, thus disentangling the underlying factors of variation.\par
Following the disentangled learning, we fuse the disentangled representations of the two modalities via a feed-forward network composed of one layer. The fused representation $h_f$ is given by,
\begin{equation}\label{eq:fused_representation}
    h_f = f(WN([h_t; h_v]) + b),
\end{equation}
where $W$ and $b$ are the weights and bias of the fusion layer respectively, [.;.] denotes the concatenation operation, $f$ the activation function and $N(.)$ the layer normalization. This fused representation $h_f$ is then used in the matrix reconstruction step, to leverage the complementary information from both textual and visual modalities.

\subsection{Cross-Attention for User's Preference Modeling}
In our framework, we employ a cross-attention mechanism to model user preferences by integrating the fused multimodal representations learned during the pre-training phase. This approach aims to capture the nuanced interactions between users and items by attending to various aspects of the multimodal features, thus improving the accuracy of rating predictions.\par
Given the fused disentangled embeddings of text and image $h_f$, we apply the cross-attention mechanism in conjunction with the rating matrix $\mathcal{R}$. Specifically, let $Q$, $K$ and $V$ represent the query, key and value matrices respectively. In our cross-attention setup, the rating matrix $\mathcal{R}$ serves as the query $Q$, while the fused multimodal representations $h_f$ forms the keys $K$ and values $V$. The cross-attention operation is defined as,
\begin{equation}\label{eq:cross-attention}
    \text{Cross-Attention}(Q,K,V) = \text{softmax}\left (\frac{QK^T}{\sqrt{d}}\right )V,
\end{equation}
where $Q, K, V\in \mathbb{R}^{I\times l}$, with $I$ being the number of items and $l$ the latent dimension of the query, key and value. This mechanism computes a weighted sum of the values $V$ based on the similarity (i.e., dot product) between the query $Q$ and the keys $K$, allowing the model to focus on the most relevant aspects of the item's multimodal features for each user. The output of the cross-attention is an intermediate representation of the user-item rating matrix. We then apply a feed-forward layer to project this representation back to its original dimension.

\subsection{Rating Matrix Reconstruction}

After generating the cross-attention-based rating matrix, we follow the approach proposed by \cite{han2021glocal} which involves refining the predictions using the AE. The rating matrix produced in the previous steps serves as input to the trained AE model, which has been pretrained to capture the underlying structure of user-item interactions. In this phase, the AE model takes its pretrained weights and fine-tunes them based on the cross-attention-modified rating matrix, as depicted in Fig. \ref{fig:cadmr_architecture}.\par

This adjustment allows the model to integrate the additional information encapsulated by the cross-attention mechanism, leading to a more accurate reconstruction of user preferences. The output of this fine-tuned AE corresponds to the final predicted ratings in the proposed recommender system. By reconstructing the rating matrix with the enhanced input, the system ensures that the final predictions are informed by both the original latent factors learned during pretraining and the nuanced multimodal interactions captured through cross-attention.

\subsection{Discussion} We identified two main reasons for choosing cross-attention in our recommendation framework: its ability to model the heterogeneity of user preferences and item characteristics, and its contribution to model interpretability.\par
From a modeling perspective, cross-attention excels at addressing the diverse and complex nature of user preferences and item attributes. Users often display varied tastes that are influenced by different factors depending on the context. For instance, a user might prioritize certain visual cues in an image when selecting a movie to watch. Similarly, items in a recommender system can be highly diverse, each possessing a unique set of attributes that appeal to different users. Cross-attention allows the model to selectively align the most relevant item features with each user's preferences. From an interpretability perspective, cross-attention mechanisms offer valuable insights into which features are most influential in driving user decisions. By explicitly modeling the interactions between user preferences and item features, cross-attention highlights the aspects of the data that contribute most to the recommendations.

\section{Training}

\subsection{Regularization}
To prevent overfitting and enhance generalization, we employ two types of regularization during training:

\paragraph{$L_2$ Regularization} Following \cite{han2021glocal}, we apply $L_2$ regularization (Ridge regularization) to both the weight and kernel matrices. This is controlled by separate penalty parameters $\lambda_2$ and $\lambda_s$.

\paragraph{Dropout} We use a dropout rate of $Dr_{rate}=0.2$ during the initial step of multimodal feature extraction. Dropout is applied to the output of each sub-layer before it is normalized and passed as input to the subsequent layer.

\subsection{Hardware and Schedule}
We trained our CADMR model on a single NVIDIA A100 GPU. With the hyperparameters discussed throughout the paper, each training step took approximately 1 second on average. For comparison, training the model on a machine with only a CPU took around 10 seconds per step.

\subsection{Optimization}
We trained our model using a squared errors loss with the aformentionned $L_2$ regularization term plus the additional Total Correlation loss for the disentangled representation learning. The final objective function of CADMR is defined as,
\begin{equation}
    \mathcal{L} = \mathcal{L}_{MSE} + \lambda \mathcal{L}_{TC},
\end{equation}
where $\mathcal{L}_{MSE}$ and $\mathcal{L}_{TC}$ are the mean squared and total correlation losses respectively. $\lambda$ is a hyperparameter that controls the relative importance of the disentangled representation learning total correlation loss with respect to the mean squared loss, which is set to $0.5$.\par
We utilized the Adam optimizer \cite{kingma2014adam} with $\beta_1=0.9$, $\beta_2=0.99$, and $\epsilon=10^{-6}$ to ensure efficient and stable convergence.

\section{Experiments}
To validate the effectiveness of our proposed multimodal recommender system, we conducted extensive experiments on three benchmark datasets. In what follows, we introduce the experimental setup and then report and analyze the experimental results.
\subsection{Experimental Setup}
\subsubsection{Datasets}
The Amazon products dataset \cite{hou2024bridging} is a widely-used benchmark in the field of multimodal recommender systems, providing a rich and diverse set of information for model training and evaluation. It is divided into 24 product categories and contains 48 million items, and 571 million reviews from 54 million users represented as reviews; i.e., ratings, text, helpfulness votes. Additionally most of the items are accompanied by the corresponding metadata; i.e., descriptions, category information, price, brand, and image features. For our experimentations, three product subsets are selected: Baby, Sports and Electronics. Table \ref{tab:datasets} presents the general information about each of the selected products subsets.\par
Following \cite{zhou2023mmrec,wei2020graph,wei2019mmgcn}, we reduce the Amazon datasets to the 5-core setting, that is each of the users and items have 5 reviews each. The resulting datasets are then randomly split into training, validation and test sets using the ratio 8:1:1.\par
The available metadata is used as our multimodal information. Specifically, for the text modality, we concatenate the item's title, description, brand, and categories. We then apply a pretrained SBERT model \cite{reimers2019sentence} to extract the textual features, resulting in a 384-dimensional vector. Visual features from each item's image are extracted using a Deep CNN, producing a 4096-dimensional vector as presented by \cite{zhang2021mining}. Finally, we construct the rating matrix based on the user-item interactions.\par

\begin{table}[t]
\centering
\caption{Datasets used in our experiments}
\begin{tabular}{@{}lccc@{}}
\toprule
\textbf{Dataset} & \textbf{\# Users} & \textbf{\# Items} & \textbf{\# Interactions} \\ \midrule
Baby             & 19445             & 7050              & 160,792                  \\ \midrule
Sports           & 35598             & 18357             & 296,337                  \\ \midrule
Electronics      & 192,403           & 63,001            & 1,689188                 \\ \bottomrule
\end{tabular}
\label{tab:datasets}
\end{table}

\subsubsection{Metrics}
There are several widely-used evaluation metrics for recommendation tasks, including accuracy, recall, precision, Normalized Discounted Cumulative Gain (NDCG), and hit rate. These metrics are commonly used for evaluating recommender systems, where higher values signify better performance in identifying relevant items. Given that the rating matrix represents user-item interactions, we define an interaction as $r_{ij}=1$ (positive) if a user has interacted with an item, and $r_{ij}=0$ (negative) otherwise. Consequently, the dataset contains both positive (interaction) and negative (no interaction) labels.

Due to the typically high sparsity of recommendation datasets, often exceeding $90\%$, the dataset is inherently unbalanced, with the majority of labels being negative. In such cases, accuracy can be misleading, as a model could achieve a high accuracy score by predominantly predicting negative labels. To address this imbalance, we focus primarily on two metrics in our experiments: Recall and NDCG, specifically their variants Recall@K and NDCG@K, where $K \in \{10, 20\}$.
\paragraph{Recall$@K$} measures the proportion of correctly predicted relevant items within the top $K$ recommendations relative to the total number of relevant items in the dataset. However, this metric does not account for the order or relevance of the items within the list.

\paragraph{NDCG$@K$} (Normalized Discounted Cumulative Gain) evaluates the quality of the ranking by considering the relevance of the recommended items and their positions in the list. It computes the relevance score of items in the recommendation list and discounts it logarithmically, penalizing relevant items that are not ranked at the top. The ideal ordered recommendation list (iDCG) serves as the ground truth for this metric. NDCG@K is defined as: \begin{equation}\label{eq
} \text{NDCG@K} = \frac{1}{\text{iDCG}}\sum_{k=1}^K\frac{2^{\text{relevance}_k} - 1}{\log _2(k+1)} \end{equation}

\subsubsection{Compared methods}
We compare CADMR with seven multimodal recommender systems baselines, namely,
\begin{itemize}
    \item LATTICE \cite{zhang2021mining} introduces a modality-aware structure learning layer that generates item-item structures for each modality and combines them to create latent item graphs. Graph convolutions on these graphs explicitly embed high-order item affinities into item representations, which can be integrated into existing collaborative filtering models to enhance recommendation accuracy.
    \item BM3 \cite{zhou2023bootstrap} begins by bootstrapping latent contrastive views from user and item representations using simple dropout augmentation. It then simultaneously optimizes three multimodal objectives to learn user and item representations by reconstructing the user-item interaction graph and aligning modality features across both inter- and intra-modality perspectives.
    \item The two primary components of SLMRec \cite{tao2022self} are (1) data augmentation of multimodal content, which generates multiple item views using three operators: feature dropout (FD), feature masking (FM), and fine and coarse feature spaces (FAC); and (2) contrastive learning, which separates an item's view from others' views to provide extra supervisory signals.
    \item ADDVAE \cite{tran2022aligning} introduces a second set of disentangled user representations learned from textual content and aligns these with the original set, improving both recommendation effectiveness and representation interpretability.
    \item FREEDOM \cite{zhou2023tale} proposes a straightforward yet effective model that freezes the item-item graph while simultaneously denoising the user-item interaction graph to enhance multimodal recommendations.
    \item DRAGON \cite{zhou2023enhancing} builds a user-user graph based on common item interactions and an item-item graph from multimodal item features. It uses graph learning on both the heterogeneous user-item graph and the homogeneous user-user and item-item graphs to derive dual user and item representations. Finally, an attentive concatenation method is utilized for fusing user and item representations.
    \item MG \cite{zhong2024mirror} introduces Mirror Gradient (MG), a concise yet effective gradient strategy that enhances model robustness during optimization, reducing instability risks from multimodal inputs.
\end{itemize}

\begin{table}[]
\centering
\caption{Performance comparison by different multimodal recommender system models in terms of NDCG and Recall. The best results are in bold.}

\begin{tabular}{@{}cccccc@{}}
\toprule
\multicolumn{1}{l}{Dataset} & Model       & NDCG@10               & NDCG@20               & \multicolumn{1}{l}{Recall@10} & \multicolumn{1}{l}{Recall@20} \\ \midrule
\multirow{8}{*}{Baby}       & LATTICE     & 0.0292                & 0.0370                & 0.0547                        & 0.0850                        \\ \cmidrule(l){2-6} 
                            & BM3         & 0.0301                & 0.0383                & 0.0564                        & 0.0883                        \\ \cmidrule(l){2-6} 
                            & SLMRec      & 0.0290                & 0.0353                & 0.0529                        & 0.0775                        \\ \cmidrule(l){2-6} 
                            & ADDVAE      & 0.0323                & 0.0404                & 0.0598                        & 0.091                         \\ \cmidrule(l){2-6} 
                            & FREEDOM     & 0.0330                & 0.0424                & 0.0627                        & 0.0992                        \\ \cmidrule(l){2-6} 
                            & DRAGON      & 0.0345                & 0.0435                & 0.0662                        & 0.1021                        \\ \cmidrule(l){2-6} 
                            & DRAGON + MG & 0.0369                & 0.0465                & 0.0701                        & 0.1082                        \\ \cmidrule(l){2-6} 
                            & CADMR       & \textbf{0.1693}       & \textbf{0.1779}       & \textbf{0.2640}               & \textbf{0.2940}               \\ \midrule
\multirow{8}{*}{Sports}     & LATTICE     & 0.0335                & 0.0421                & 0.0620                        & 0.0953                        \\ \cmidrule(l){2-6} 
                            & BM3         & 0.0355                & 0.0438                & 0.0656                        & 0.0980                        \\ \cmidrule(l){2-6} 
                            & SLMRec      & 0.0365                & 0.0450                & 0.0663                        & 0.0990                        \\ \cmidrule(l){2-6} 
                            & ADDVAE      & \textit{0.0389}       & \textit{0.0473}       & \textit{0.0709}               & \textit{0.1035}               \\ \cmidrule(l){2-6} 
                            & FREEDOM     & 0.0385                & 0.0481                & 0.0717                        & 0.1089                        \\ \cmidrule(l){2-6} 
                            & DRAGON      & 0.0403                & 0.0500                & 0.0749                        & 0.1124                        \\ \cmidrule(l){2-6} 
                            & DRAGON + MG & 0.0431                & 0.0535                & 0.0793                        & 0.1191                        \\ \cmidrule(l){2-6} 
                            & CADMR       & \textbf{0.1719}       & \textbf{0.1796}       & \textbf{0.2754}               & \textbf{0.2977}               \\ \midrule
\multirow{8}{*}{Elec.}      & LATTICE     & \multicolumn{1}{l}{-} & \multicolumn{1}{l}{-} & \multicolumn{1}{l}{-}         & \multicolumn{1}{l}{-}         \\ \cmidrule(l){2-6} 
                            & BM3         & 0.0247                & 0.0302                & 0.0437                        & 0.0648                        \\ \cmidrule(l){2-6} 
                            & SLMRec      & \textit{0.0249}       & \textit{0.0303}       & \textit{0.0443}               & \textit{0.0651}               \\ \cmidrule(l){2-6} 
                            & ADDVAE      & 0.0253                & 0.0308                & 0.0451                        & 0.0665                        \\ \cmidrule(l){2-6} 
                            & FREEDOM     & 0.0220                & 0.0273                & 0.0396                        & 0.0601                        \\ \cmidrule(l){2-6} 
                            & DRAGON      & 0.0292                & 0.0357                & 0.0522                        & 0.0781                        \\ \cmidrule(l){2-6} 
                            & DRAGON + MG & 0.0312                & 0.0381                & 0.0553                        & 0.0827                        \\ \cmidrule(l){2-6} 
                            & CADMR       & \textbf{0.1245}       & \textbf{0.1310}       & \textbf{0.2253}               & \textbf{0.2489}               \\ \bottomrule
\end{tabular}
\label{tab:quantitative_results}
\end{table}

\subsection{Performance Comparison Analysis}
Our proposed model, CADMR, demonstrates a substantial improvement over state-of-the-art multimodal recommender systems across key metrics, including NDCG$@10$, NDCG$@20$, Recall$@10$, and Recall$@20$, as demonstrated in Table \ref{tab:quantitative_results}. Specifically, CADMR significantly outperforms competing models in terms of both NDCG and Recall. For instance, in the context of NDCG$@10$, CADMR achieves a notable improvement, with values such as $0.1693$, $0.1719$, and $0.1245$ across the evaluated datasets, far surpassing the closest competitor, MG, which reaches 0.0431 in the best case.\par
Similarly, for Recall$@10$, it attains values such as 0.2640, 0.2754, and 0.2253, which are more than three times higher than the best-performing baseline model. This increase in recall indicates CADMR's proficiency in retrieving a higher proportion of relevant items within the top recommendations, thus enhancing the user experience by presenting more relevant suggestions.

The consistent outperformance across various datasets and metrics demonstrates CADMR's generalizability and efficiency in leveraging multimodal information to refine recommendations. This success can be attributed to the novel integration of cross-attention mechanisms, which enable the model to capture intricate interactions between users and items, leading to more accurate and personalized recommendations. Overall, CADMR sets a new benchmark in multimodal recommendation, significantly advancing the state of the art.

\subsection{Cold start Analysis}
\begin{figure}
    \centering
    \caption{CADMR performance with respect to the training set size.}
    \includegraphics[width=0.5\linewidth]{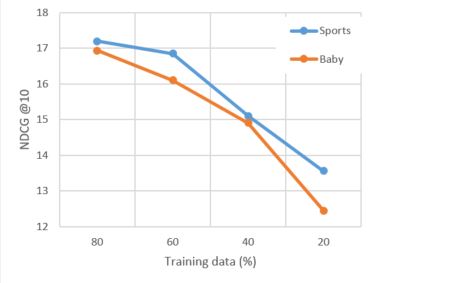}
    \label{fig:cold_start}
\end{figure}

We wanted to evaluate the model's resilience to the cold-start problem by varying the size of the training data from 80\% to 20\%. The results, depicted in Fig. \ref{fig:cold_start}, show the NDCG$@10$ performance on the Sports and Baby datasets.\par

As expected, the NDCG@10 score decreases as the training data size diminishes, which is a common manifestation of the cold-start problem. The reduction in available training data limits the model's ability to learn user-item interactions effectively, leading to a decline in recommendation quality. For the Sports dataset, the NDCG$@10$ progressively drops to approximately $13\%$ when the training data is reduced to 20\%. Similarly, the Baby dataset follows a similar outcome, with NDCG$@10$ decreasing to just below $13\%$ over the same range. \par

Despite this decrease in performance, it is important to note that our model, CADMR, still outperforms existing methods across all training sizes. Even at the lowest training size of 20\%, CADMR maintains a higher NDCG$@10$ than what many competing models achieve with a full dataset. This demonstrates the robustness of CADMR in handling sparse data scenarios and mitigating the cold-start problem better than prior state-of-the-art models.
These results underline CADMR's strong generalization capability, making it a competitive solution even in data-constrained environments.

\subsection{Ablation Study}
In this section, we present an ablation study to evaluate the contribution of two key components in our model, CADMR: the cross-attention mechanism and the disentangled representation learning (DRL) module. By systematically removing these components, we measure their individual impact on the overall performance. We compare the following variants:
\begin{itemize}
    \item $CADMR_{\text{w/o}\hspace{0.1cm} \text{DRL}}$: This variant removes the disentangled representation learning module, retaining only the cross-attention mechanism.
    \item $CADMR_{\text{w/o}\hspace{0.1cm} \text{CA}}$: In this version, we exclude the cross-attention mechanism while keeping the disentangled learning module.
\end{itemize}

The results of the ablation study are summarized in Table \ref{tab:ablation_study}. It is evident that removing either of these components leads to a significant decline in performance across all datasets (Baby, Sports, and Electronics). Specifically, when DRL is removed, the performance drops notably, with NDCG@10 scores of 0.1215 on the Baby dataset, 0.1376 on the Sports dataset, and 0.1083 on the Electronics dataset. Similarly, excluding the cross-attention mechanism ($\text{CADMR}_{w/o\hspace{0.1cm}CA}$) results in reduced performance, with NDCG@10 scores of 0.0639,  0.0681, and 0.0527 for Baby, Sports, and Electronics datasets, respectively. This demonstrates the substantial contribution of cross-attention to our model's ability to enhance user-item interaction with multimodal data effectively. Our base CADMR model, which integrates both cross-attention and disentangled representation learning, achieves the highest scores on all datasets. This confirms that both components are essential for maximizing performance, with their combined effect yielding the best results. 

\begin{table}[t]
\centering
\caption{Performance of the different variants of our CADMR model over all three datasets. Best results are in bold.}
\begin{tabular}{@{}lccc@{}}
\toprule
Method                        & Baby           & Sports         & Elec.          \\ \midrule
$CADMR_{\text{w/o}\hspace{0.1cm}\text{DRL}}$   & 0.1215         & 0.1376         & 0.1083         \\ 
$CADMR_{\text{w/o}\hspace{0.1cm}\text{CA}}$    & 0.0639         & 0.0681         & 0.0527         \\ 
$CADMR_{\text{base}}$                          & \textbf{0.1693} & \textbf{0.1719} & \textbf{0.1245} \\ \bottomrule
\end{tabular}
\label{tab:ablation_study}
\end{table}

\subsection{Effect of Cross-Attention heads number}
We varied the number of cross-attention heads to assess their effect on our model’s performance across the Baby, Sports, and Electronics datasets. As the number of heads increased from 1 to 4, a clear improvement in NDCG$@10$ scores was observed in all datasets. This steady improvement indicates that adding more cross-attention heads up to 4 significantly enhances the model’s ability to capture complex multimodal relationships.\par
However, when the number of heads was increased beyond 4, the performance gains began to taper off. On all three datasets, moving from 4 to 8 heads resulted in only marginal improvements, signaling a plateau in performance. These results suggest that while cross-attention heads are beneficial up to a certain point, adding more heads beyond 4 yields diminishing results.

\begin{figure}[t]
    \centering
    \includegraphics[width=\linewidth]{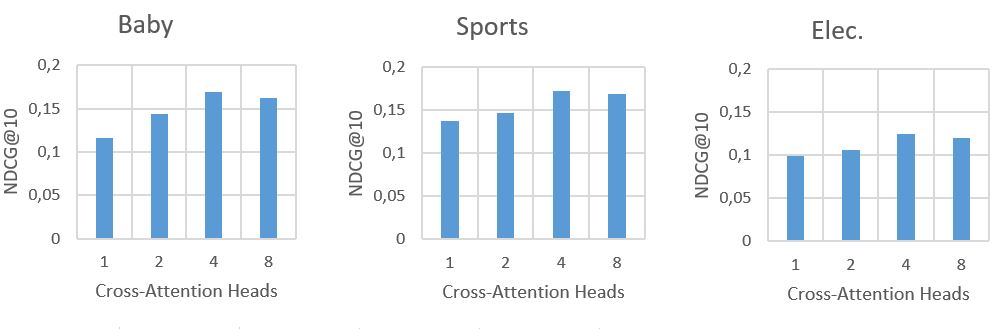}
    \caption{Impact of the cross-attention number of heads on the overall performance of our model}
    \label{fig:n_head}
\end{figure}

\section{Conclusion}
In this paper, we introduced CADMR, an autoencoder-based multimodal recommender system designed to address the challenges of high-dimensional, sparse user-item rating matrices. CADMR integrates multi-head cross-attention mechanisms and disentangled learning to effectively leverage diverse multimodal data, enhancing both the reconstruction of the rating matrix and the representation of user-item interactions. By disentangling modality-specific features while preserving their interdependence, CADMR learns a joint latent representation that significantly improves recommendation performance. Extensive evaluations on multiple benchmark datasets demonstrated its superior performance over state-of-the-art methods, as well as the critical role of the cross-attention and disentangled learning modules. In future work, we will further analyze CADMR’s scalability and explore potential avenues for enhancing its effectiveness.

\newpage



\end{document}